\documentclass[twocolumn,showpacs,preprintnumbers,amsmath,amssymb,prl,superscriptaddress]{revtex4}

\usepackage{amsmath,amssymb,amsthm,color}
\usepackage{graphicx}
\usepackage{dcolumn}

\newcommand {\beq}{\begin{eqnarray}}
\newcommand {\eeq}{\end{eqnarray}}

\begin{document}

\preprint{IPMU11-0163}

\title{Comments on scale invariant but non-conformal supersymmetric field theories}

\author{Yu Nakayama}

\affiliation{Institute for the Physics and Mathematics of the Universe,  \\ Todai Institutes for Advanced Study,
University of Tokyo, \\ 
5-1-5 Kashiwanoha, Kashiwa, Chiba 277-8583, Japan}

\affiliation{California Institute of Technology, 452-48, Pasadena, California 91125, USA}


\begin{abstract}
We investigate a possibility of scale invariant but non-conformal supersymmetric field theories from a perturbative approach. The explicit existence of 
monotonically decreasing $a$-function that generates beta-functions as a gradient flow provides a strong obstruction for such a possibility at two-loop order. We comment on the ``discovery" of scale invariant but non-conformal renormalization group trajectories via a ``change of scheme" in $(4-\epsilon)$ dimension proposed in literatures.
\end{abstract}

\maketitle

\section{Introduction}

It is a long-standing problem whether scale invariant unitary relativistic quantum field theories in four dimension actually show conformal invariance. In two dimension, the equivalence was proved \cite{Zamolodchikov:1986gt}\cite{Polchinski:1987dy}\cite{Mack1}, and in higher dimension with $d>4$, there is at least one explicit counterexample \cite{Jackiw:2011vz}\cite{ElShowk:2011gz}. A search for counterexamples in $d = 4-\epsilon$ has been discussed \cite{Dorigoni:2009ra}\cite{Fortin:2011ks}\cite{Fortin:2011sz}. We also have a gravitational argument supporting the equivalence \cite{Nakayama:2009qu}\cite{Nakayama:2009fe}\cite{Nakayama:2010wx}.

In this paper, we study a possibility to construct scale invariant but non-conformal field theories with a help of supersymmetry in four as well as $(4-\epsilon)$ dimension. The general structure was investigated in \cite{Antoniadis:2011gn}, but without the detail of the beta-functions, it has been unknown whether the scale invariant supersymmetric field theories are superconformal or not.
We show that the existence of monotonically decreasing $a$-function in supersymmetric field theories at two-loop order hinders scale invariant but non-conformal renormalization group trajectories within a perturbation theory. Thus, at two-loop order, we conclude that supersymmetric scale invariant field theories are all superconformal.

We also comment on the ``discovery" of scale invariant trajectories via a ``change of scheme" in $(4-\epsilon)$ dimension proposed by \cite{Fortin:2011ks}\cite{Fortin:2011sz}. We argue that the existence of the scale invariant but non-conformal trajectories as well as the monotonically decreasing $a$-function indeed depend on their ``change of scheme" but such a change is not natural at least in supersymmetric field theories. This subtlety is intrinsic to the ambiguity in $\epsilon$ expansion and should be absent in strict four-dimensional theories.

\section{Formalism}

In scale invariant relativistic quantum field theories in arbitrary dimension greater than two, trace of an energy-momentum tensor $T_{\mu\nu}$ is given by the divergence of a Virial current $V_\mu$ \cite{Coleman:1970je}:
\begin{align}
T^{\mu}_{\ \mu} = \partial^\mu V_\mu \ .
\end{align}
The theory is conformal invariant if and only if we can improve the energy-momentum tensor so that it is traceless. In terms of the Virial current, if and only if the Virial current is a total derivative 
\begin{align}
V_\mu = \partial^\nu L_{\nu \mu} + J_\mu
\end{align}
up to a conserved current $J_\mu$ (i.e. $\partial^\mu J_\mu = 0$), we can improve the energy-momentum tensor so that the theory is conformal \cite{Polchinski:1987dy}.

In this paper, we consider the supersymmetric field theories in $(4-\epsilon)$ dimension with R-symmetry that allow perturbative scale invariance. Such theories consist of vector multiplets whose field strength supermultiplets are $W_i$ with arbitrary gauge group and their coupling constants $g^2_{ij}$, and matter chiral multiplets $\Phi_a$ with arbitrary representation (modulo anomaly cancellation) and Yukawa-coupling constants $Y_{abc}$. The theory is completely specified by the superpotential
\begin{align}
W = \frac{1}{g_{ij}^2} W_i W_j + Y_{abc} \Phi^a \Phi^b \Phi^c \ .
\end{align}
Note that Yukawa-coupling constants $Y_{abc}$ can be chosen to be completely symmetric in their indices. One may introduce $\theta$-parameters, but they will not affect the following perturbative analysis.

It was shown \cite{Antoniadis:2011gn} that the structure of the Virial current supermultiplet $\mathcal{V}_\mu$, whose lowest component is $V_\mu$, is highly constrained from the unitarity in supersymmetric field theories with R-symmetry:
\begin{align}
\mathcal{V}_\mu = \partial_\mu U + \sigma_{\mu}^{\dot{\alpha} \beta} [\bar{D}_{\dot{\alpha}}, D_\beta] \mathcal{O} \ .
\end{align}
Here $U$ and $\mathcal{O}$ are dimension-two singlet real scalar operators.
The first term is irrelevant for our study because we can always improve the energy momentum tensor to eliminate it.

To all orders in perturbation theory, $\mathcal{O}$ must take the following form:
\begin{align}
\mathcal{O} = Q_{b}^{\ a} \bar{\Phi}_a \Phi^b \ , \label{virial}
\end{align}
where $Q_{b}^{\ a}$ is an Hermitian matrix. Actually, the same conclusion follows even if we did not assume the R-symmetry \footnote{The discussion in \cite{Zheng:2011bp} is incomplete, and the relation between the dilatation multiplet and the Virial current multiplet in non R-symmetric case must be modified with the additional $x^{\mu}(D^2 {X}+\bar{D}^2\bar{X})$ term in their formula. After the suitable modification, one can show that the Virial current multiplet is given by $\mathcal{V}_\mu = \sigma^{\alpha \dot{\alpha}}_\mu(\bar{D}_{\dot{\alpha}} \Gamma_{\alpha} - D_{\alpha} \bar{\Gamma}_{\dot{\alpha}})$. Within a perturbation theory, \eqref{virial} is anyway the only possibility with $\Gamma_{\alpha} = D_{\alpha}(\bar{\Phi} \Phi)$. Besides, the one-loop fixed point is superconformal, so the theory should possess R-symmetry.}.
The corresponding Virial current is
\begin{align}
V_\mu = iQ_{b}^{\ a} (\partial_\mu \bar{\phi}_a  \phi^b - \bar{\phi}_a \partial_\mu \phi^b + \bar{\psi}_a \gamma_\mu \psi^b) . 
\end{align}
The application of the equation of motion gives the divergence of the Virial current
\begin{align}
&\partial^\mu V_\mu \cr
&= i (Q_{a}^{\ d} Y_{d bc} + Q_{b}^{\ d}Y_{adc} + Q_{c}^{\ d} Y_{abd}) (\phi^{a} \psi^{b} \psi^c)  + c.c. \cr
 &+ Q|Y|^2\phi^4 .
\end{align}
The supersymmetry relates the Yukawa-interaction with the $\phi^4$ interaction schematically represented by $Q|Y|^2 \phi^4$,  and the detail of the latter is not important in our study. Here it is important to note that we have implicitly assumed the existence of the renormalization scheme where the equation of motion holds in the perturbation theory.

On the other hand, the trace of the energy-momentum tensor is given by the beta-functions of the coupling constants $g^2_{ij}$ and $Y^{abc}$.
\begin{align}
T^{\mu}_{\ \mu} &= \beta^{g^{2}_{ij}} F_{\mu \nu}^i F^{j \mu\nu} \cr
 &+\beta^{Y_{abc}} (\phi^{a } \psi^{b} \psi^c) + c.c. \cr 
&+ \cdots \ ,
\end{align}
where $\cdots$ denotes the other terms that complete the supersymmetry such as $\phi^4$ interactions or those that involve gauginos. Due to the supersymmetry, these omitted terms will not add any new information in the following discussion.

Scale invariance means that the trace of the energy-momentum tensor must be expressed as a divergence of the Virial current. 
First of all, when $\epsilon = 0$, it demands that the beta-functions for the gauge coupling constant must vanish
\begin{align}
\beta^{{g}^{2}_{ij}} = 0 \ . \label{beta1}
\end{align}
We assume that these equations are solved.

When $\epsilon = 0$, the additional condition from the Yukawa-coupling constants is
\begin{align}
 &i (Q_{a}^{\ d} Y_{d bc} + Q_{b}^{\ d}Y_{adc} + Q_{c}^{\ d} Y_{abd}) \cr
&= (\Gamma_{a}^{\ d} Y_{d bc} + \Gamma_{b}^{\ d}Y_{adc} + \Gamma_{c}^{\ d} Y_{abd}) \ . \label{beta2}
\end{align}
Here, we have used the fact that the anomalous dimension $\Gamma_{a}^{\ b}$ will determine the beta-functions of the Yukawa-coupling constants in supersymmetric theories.
A priori, it is not obvious whether all the solutions of  \eqref{beta1} and \eqref{beta2} demand $QY \equiv (Q_{a}^{\ d} Y_{d bc} + Q_{b}^{\ d}Y_{adc} + Q_{c}^{\ d} Y_{abd}) = 0 $. In $\mathcal{N}=2$ supersymmetric case, \eqref{beta1} means that the anomalous dimension matrix $\Gamma_{a}^{\ b}$ automatically vanish, so obviously they are superconformal. In generic $\mathcal{N}=1$ supersymmetric case, we need more detailed information of the beta-functions. One of the advantages of the supersymmetry, however, is that the stability of the potential is automatically guaranteed unlike the non-supersymmetric case studied in \cite{Fortin:2011ks}\cite{Fortin:2011sz}.

At one-loop order,  $\Gamma_{a}^{\ d} = \frac{1}{2}Y_{abc} \bar{Y}^{dbc} + 3g^2 (t^2)^{\ d}_{a}$ \footnote{In this paper, we use the convention that our coupling constants are divided by $(4\pi)$ compared with the standard textbook convention.}, 
so by contracting $QY$ on the both side of \eqref{beta2}, and by observing that the left-hand side is a pure imaginary number while the right-hand side is a real number, we conclude $Q Y$ must vanish at this order (see \cite{Dorigoni:2009ra}\cite{Fortin:2011ks} for a similar argument in non-supersymmetric case). At two-loop or higher, we have not  been aware of a simple proof of vanishing of $QY$. From phenomenological studies with a small number of matters, however, all the solutions of \eqref{beta2} seem to indicate $QY =0$ via analytical as well as numerical approach.
In the next section, we will give an indirect evidence of the non-existence of scale invariant but non-conformal trajectories in these perturbative supersymmetric theories based on the explicit existence of the monotonically decreasing $a$-function that generates beta-functions as a  gradient flow.

\section{Perturbative existence of $a$-function}
In this section, we embody the idea to utilize the existence of monotonically decreasing $a$-function \cite{Freedman:1998rd} to show the non-existence of scale invariant but non-conformal field theories at two-loop order.
Let us consider the most generic Wess-Zumino model in $(4-\epsilon)$ dimension, which is completely specified by the Yukawa-coupling constants $Y_{abc}$. We can directly show that the two-loop beta-function \cite{Jack:1996qq}
\begin{align}
\beta_{Y_{abc}} = -\epsilon Y_{abc} +  (\Gamma_{a}^{\ d} Y_{d bc} + \Gamma_{b}^{\ d}Y_{adc} + \Gamma_{c}^{\ d} Y_{abd}) \label{betatwo} 
\end{align}
with the anomalous dimension matrix
\begin{align}
\Gamma_{a}^{\ b} = P_{a}^{\ b} -\bar{Y}^{b cd} Y_{a ce} P_d^{\ e} \ ,
\end{align}
where $P_{a}^{\ b} = \frac{1}{2} \bar{Y}^{bcd} Y_{acd}$, 
can be generated by an ``$a$-function"
\begin{align}
a &= a_0 - \frac{\epsilon}{3} \mathrm{tr}(\bar{Y}Y)  \cr
 &+ (\frac{1}{4} + \frac{\epsilon}{8}) \mathrm{tr}((\bar{Y}Y)(\bar{Y}Y))- \bar{Y}^{acd}Y_{bce}P^{\ e}_{d}P^{\ b}_{a} \cr
 &- \frac{1}{24}\mathrm{tr}((\bar{Y}Y)(\bar{Y}Y)(\bar{Y}Y))  \ , \label{a}
\end{align}
where $a_0$ is a constant and $(\bar{Y} Y)^{\ a}_{b} = \bar{Y}^{acd}Y_{bcd}$, with the ``metric" of the coupling constant space (symbolically denoted by $g^I$)
\begin{align}
ds^2 &= G_{IJ} dg^{I}dg^{J} \cr
 &= \frac{2}{3} d\bar{Y}^{abc} dY_{abc} -\frac{1}{2} \mathrm{tr}((d\bar{Y} dY)(\bar{Y} Y) ) \ . 
\end{align}
as a gradient flow
\begin{align}
\partial_{Y_{abc}} a = G_{Y_{abc} \bar{Y}^{efg}}\beta^{\bar{Y}^{efg}} \ . 
\end{align}
We have added $\epsilon$ dependent terms to the result found in the literature \cite{Freedman:1998rd} in order to reproduce the classical part in the beta-function \eqref{betatwo}. Up to two-loop accuracy, we still have some freedom to modify the $a$-function as well as the metric \cite{Freedman:1998rd}, but it is not relevant for our study of two-loop search of the scale invariant trajectories. We also note that our $a$ may be different from the actual ``central charge" away from the conformal invariant fixed point.

This gradient flow immediately implies that the $a$-function is monotonically decreasing
\begin{align}
\frac{da}{d\log \mu} = - \beta^{Y_{abc}} G_{Y_{abc} \bar{Y}^{efg}} \beta^{\bar{Y}^{efg}}
\end{align}
for sufficiently small coupling constants $Y_{abc} \sim \sqrt{\epsilon} \ll 0$ near all one-loop conformal fixed points because the ``metric" is positive definite when $\epsilon \ll 1$.

We now argue that the existence of scale invariant but non-conformal trajectories at two-loop is inconsistent with the monotonically decreasing $a$-function explicitly constructed here. Suppose that there exists such a trajectory, then along the trajectory, $a$ continues to decrease:
\begin{align}
\frac{da}{d\log \mu} =- (\bar{Y} Q) G (Q Y) < 0 \ .
\end{align}

At this point, we recall that the two-loop scale invariant trajectory should approximately lie on the one-loop conformal fixed points. The one-loop conformal fixed points form a trajectory generated by the action of $Q$ on a given reference one-loop conformal fixed point $Y_0$ (which is nothing but the field redefinition to generate new fixed points), and the two-loop scale invariant trajectory, if any, approximately moves along this trajectory generated by the same $Q$ with the speed of order $\epsilon^2$ for its one-loop consistency \footnote{The one-loop fixed point would never nontrivially ``split" into more than two two-loop trajectories with $Q=0$ and $Q\neq 0$. If this could happen, we would construct interpolating solutions between $Q=0$ and non-zero $Q$ at a fixed order of perturbative expansion by using the linearity of the perturbation equation. Since we could construct arbitrary large $Q$ solution in this way, it would be reasonable only when the Virial current corresponding to $Q$ is conserved.}.

In particular, after a sufficiently large time (at least of order $\epsilon^{-2}$), the two-loop scale invariant trajectory generated by non-zero beta-functions will return to the original point $Y_0$ (when the trajectory closes), or it comes back arbitrary close to the original point (when the trajectory is chaotic) \cite{Fortin:2011ks}\cite{Fortin:2011sz}, while $a$ decreases forever. 
However, we know that $a$ is a globally defined smooth function on the coupling constant space, which is explicitly given by \eqref{a}, so it is impossible for $a$ to decrease perpetually along the closed or chaotic but bounded trajectory.

Thus, the existence of the explicit $a$-function at two-loop order means that we cannot find scale invariant but non-conformal trajectories in $(4-\epsilon)$ dimensional Wess-Zumino model with the same accuracy of the loop approximation. Indeed analytic as well as numerical study of the model with small numbers of fields verify this statement.  

Note that our argument is quite generic so it 
applies not only to the Wess-Zumino models in $(4-\epsilon)$ dimension, but also to supersymmetric gauge theory with arbitrary matters and Yukawa-coupling constants in strict four-dimensional limit thanks to the existence of the explicit $a$-function  \cite{Freedman:1998rd} that generates beta-functions as a gradient flow.  Thus within a perturbatively accessible regime, where the $\epsilon$ parameter is replaced by a small control parameter such as $\frac{3N_c-N_f}{N_c}$ in SQCD, it is impossible to find scale invariant but non-conformal trajectories in supersymmetric gauge theories at two-loop order.
 The existence of the gradient flow is argued positively at higher-loop orders \cite{Freedman:1998rd}\cite{Jack:1996qq}\cite{Wallace:1974dy}, and if this is true, we will never find a scale invariant but non-conformal trajectory in these models.
\section{Comment on \cite{Fortin:2011ks}\cite{Fortin:2011sz}}
Most of the discussion in this paper is actually applicable to non-supersymmetric field theories as well when we accept the existence of the monotonically decreasing ``$a$-function" that generates the renormalization group as a gradient flow. The existence was claimed to all orders in perturbation theory in \cite{Wallace:1974dy}\cite{Jack:1990eb}\cite{Freedman:1998rd}.  In particular, the scale invariant but non-conformal field theories cannot exist due to the same reason presented in the last section.

How is this consistent with the ``discovery" of the scale invariant but non-conformal renormalization group trajectory in non-supersymmetric field theories in ($4-\epsilon$) dimension proposed in \cite{Fortin:2011ks}\cite{Fortin:2011sz} at the ``two-loop" order? The authors are aware of the obstruction \cite{Wallace:1974dy}\cite{Jack:1990eb} and claim that the gradient flow does not exist in their examples. Let us elaborate on this point.

First, we emphasize that in the dimensional regularization scheme what they found actually is that all the theories under investigation are conformal invariant up to two loops where their approximation can be trusted. At this point, it must be consistent with the existence of $a$-function at  two-loop order.
Then, they proposed a most generic ``change of scheme" to show the existence of scale invariant but non-conformal trajectories at two-loop order.

We would like to clarify the nature of their ``change of scheme" here: in particular,
 it is curious how the ``change of scheme" can affect the physical conclusion whether the theory is conformal or only scale invariant. Indeed, if the ``change of scheme" used in \cite{Fortin:2011ks}\cite{Fortin:2011sz} were a mere reparametrization in the space of coupling constants, then the conclusion should not change. For instance, the condition of the conformal invariance 
\begin{align}
\beta^I = \frac{d g^{I}}{d\log\mu} = 0 
\end{align}
is invariant under the small reparametrization $ g^I \to g^I + A^{I}_{JK} g^J g^K$. So is the existence of the $a$-function:
\begin{align}
 \partial_I a = g_{IJ} \beta^{J} \ , \ \ 
\frac{da}{d\log\mu} = -g_{IJ} \beta^I \beta^J \ .
\end{align}
These equations are manifestly covariant under the reparametrization, and the physical conclusion should not depend on the change of scheme.

The resolution of the apparent puzzle is that the ``change of scheme" used in \cite{Fortin:2011ks}\cite{Fortin:2011sz} in ($4-\epsilon$) dimension is {\it not} a simple reparametrization. Suppose that in one particular dimensional regularization scheme, we have obtained the beta-function
\begin{align}
\beta^I = -\epsilon g^I + \beta^{(1) I}_{KL} g^{K}g^{L} + \beta^{(2) I}_{KLM} g^{K} g^{L} g^{M} + \cdots \ . \label{genbeta}
\end{align}
A small reparametrization $g^I \to g^I + A^{I}_{JK} g^{J}g^{K}$ will generically introduce not only the ``two-loop" shift mentioned in \cite{Fortin:2011ks}\cite{Fortin:2011sz} but also the ``one-loop" term that explicitly depends on $\epsilon$:
\begin{align}
\delta \beta^{I} = +\epsilon A^{I}_{JK} g^{J} g^{K} \ .  \label{epone}
\end{align}
This term is crucial to maintain the covariance of the conformal invariant condition as well as the existence of the perturbative $a$-function in $(4-\epsilon)$ dimension under reparametrization.

In \cite{Fortin:2011ks}\cite{Fortin:2011sz}, the explicit $\epsilon$ dependent term beyond the classical term such as \eqref{epone} is dropped by hand {\it after} the reparametrization, and this affects the structure of the conformal invariance at $O(\epsilon^2)$, which enabled them to ``discover" a scale invariant but non-conformal trajectory via the ``change of scheme". They could deliver the validity of the procedure by arguing  that they could have done the change of scheme in the strict four-dimensional limit (i.e. $\epsilon = 0$), and then added the classical piece of the beta-function so that the $\epsilon$ dependent one-loop term such as \eqref{epone} would never appear.
In other words, since the change of scheme and the introduction of non-zero $\epsilon$ do not commute, we should have known the regularization scheme where there is no other $\epsilon$ dependence in \eqref{genbeta} from the beginning to argue the existence of the $a$-function. Unless we give an independent argument which regularization is natural, we cannot conclude whether the lack of the integrability in beta-functions in $(4-\epsilon)$ dimension after the change of scheme is physical or not.

In our supersymmetric case, on the other hand, there exists a manifestly supersymmetric regularization scheme where the holomorphic structure is maintained, and the random change of scheme would not be allowed. Therefore, the existence of the perturbative $a$-function in $(4-\epsilon)$ dimension seems rather robust. 

For instance, we could try their ``change of scheme" in the general Wess-Zumino model studied in the last section. If we dropped the $\epsilon$ dependent one-loop term by hand subsequently after the small reparametrization of the Yukawa-coupling constants $Y \to A|Y|^2Y$, then we would  observe that the existence of the monotonically decreasing $a$-function is generically lost, and we would be able to find a scale invariant but non-conformal trajectories that emerge as early as at two-loop order. 

An explicit example is in order. Consider the Wess-Zumino model with four chiral fields $\Phi_i$ $i=1,2,3,4$ with the superpotential 
\begin{align}
 W = (x_1 \Phi_1 + x_2 \Phi_2) \Phi_3 \Phi_3 + (y_1 \Phi_1 + y_2\Phi_2) \Phi_4\Phi_4 \ 
\end{align}
in $(4-\epsilon)$ dimension.
We can easily show that to all orders in perturbation theory, fixed points are, if any, all conformal (e.g. $x_1 = x_2 \sim \sqrt{\epsilon}$, $y_1 = -y_2 \sim \sqrt{\epsilon}$) and the non-trivial Virial current vanishes. If we performed a ``change of scheme" induced by the reparametrization $ x_i \to x_i + k (\bar{y}_{j}y_j) y_i$ and crucially discard the $\epsilon$ dependent one-loop part of the beta-functions by hand, then the scale invariant but non-conformal trajectories emerge, where the parameter $Q$ for the Virial current $i(\bar{\Phi}_{1} \Phi^2 - \bar{\Phi}_{2} \Phi^1)$ is proportional to $k$. This change of scheme is very artificial, however, because it does not preserve the holomorphic structure as well as the spurious charge assignment for $x_i$ and $y_i$. This example clearly illustrates the possibility that the ``change of scheme" employed in  \cite{Fortin:2011ks}\cite{Fortin:2011sz} could transform even all-order conformal fixed points into a scale invariant but non-conformal trajectory at two-loop \footnote{For instance, in \cite{Fortin:2011ks}, it was argued that when the Yukawa-coupling constants can be reduced to a matrix, the Virial current must be trivial to all orders in perturbation theory due to a particular structure of the (unrenormalized) Feynman diagrams. However, it is easy to break this structure, albeit artificial, by the ``counter-terms" that are implied by the most generic ``change of scheme" discussed here.}.

\section{Conclusion}
We have investigated a possibility of scale invariant but non-conformal supersymmetric field theories from a perturbative approach. The explicit existence of 
monotonically decreasing $a$-function provides a strong obstruction for such a possibility at two-loop order. We have argued that the ``change of scheme" employed in  \cite{Fortin:2011ks}\cite{Fortin:2011sz} is not covariant under reparametrization in $(4-\epsilon)$ dimension, and this is the reason why they circumvented the obstruction coming from the existence of $a$-function in a conventional dimensional regularization scheme. 

The rather artificial lack of covariance under the reparametrization discussed in the last section due to a subtlety or ambiguity in $\epsilon$ expansion disappears  in the strict four-dimensional limit, and the existence of the $a$-function that generates beta-functions as a gradient flow should be reparametrization invariant or scheme independent. Thus, in the strict four-dimensional limit, we conclude that there indeed exists an obstruction to find a scale invariant but non-conformal field theories within a perturbative regime if the monotonically decreasing $a$-function is constructed in line with the claim in \cite{Jack:1990eb}. 
It would be interesting to see whether the non-perturbative reformulation of the $a$-function (see e.g. \cite{Barnes:2004jj}\cite{Komargodski:2011vj} for recent discussions) would lead to a  more comprehensive understanding of the relation between scale invariance and conformal invariance in four dimension. At the same time, it may be possible (beyond the perturbation theory) that the gradient flow may not exist and the scale invariant but non-conformal field theory is possible in four dimension unlike in two-dimension where the existence was proved.

The most natural value of the anomalous dimension matrix that will lead to a conformal fixed point is given by the ``$a$-maximization" procedure \cite{Intriligator:2003jj}. However, a priori, there is no guarantee that given a particular value of the anomalous dimension matrix via the $a$-maximization, we can solve the corresponding coupling constants from the actual beta-functions. One known possibility is that the corresponding value breaks the unitarity and the contribution from the accidental symmetry must be taken into account. The other logical possibility here, is that you may have to augment the Virial current contribution to solve them. Then, the (almost) derivation of the $a$-theorem would not hold. To understand the gradient flow of the $a$-function beyond the perturbation theory, the $a$-maximization with Lagrange multipliers (see e.g. \cite{Kutasov:2003ux}\cite{Erkal:2010sh}) seemed promising, but we have to overcome this inverse problem in any way.

Finally, it would be very interesting to see if the relation between the existence of the gradient flow and the scale invariant but non-conformal trajectories are clarified from the holographic viewpoint. In particular, we would like to shed more light on  the role of the null energy condition in both sides (see e.g. \cite{Nakayama:2010wx}\cite{Nakayama:2011zw}).
This direction is under investigation.

\section*{Acknowledgments}

We thank M.~Buican, S.~Rychkov and Y.~Tachikawa for stimulating discussions. We are greatly indebted to J.~-F.~Fortin for clarification of the arguments in their papers. There is no need to say, of course, that all the possible misunderstandings of their claim is on us. The work is supported by the 
World Premier International Research Center Initiative of MEXT of
Japan. 



\begin{thebibliography}{99}

\bibitem{Zamolodchikov:1986gt}
  A.~B.~Zamolodchikov,
  JETP Lett.\  {\bf 43} (1986) 730
  [Pisma Zh.\ Eksp.\ Teor.\ Fiz.\  {\bf 43} (1986) 565].

\bibitem{Polchinski:1987dy}
  J.~Polchinski,
  Nucl.\ Phys.\  B {\bf 303}, 226 (1988).


\bibitem{Mack1}
M. L\"{u}scher and G. Mack,
 1976, unpublished;

G. Mack,
``NONPERTURBATIVE QUANTUM FIELD THEORY. PROCEEDINGS, NATO
ADVANCED STUDY INSTITUTE, CARGESE, FRANCE, JULY 16-30, 1987''.

\bibitem{Jackiw:2011vz}
  R.~Jackiw, S.~-Y.~Pi,
  J.\ Phys.\ A {\bf A44}, 223001 (2011).
  [arXiv:1101.4886 [math-ph]].

\bibitem{ElShowk:2011gz}
  S.~El-Showk, Y.~Nakayama, S.~Rychkov,
  Nucl.\ Phys.\  {\bf B848}, 578-593 (2011).
  [arXiv:1101.5385 [hep-th]].

\bibitem{Dorigoni:2009ra}
  D.~Dorigoni, V.~S.~Rychkov,
  [arXiv:0910.1087 [hep-th]].

\bibitem{Fortin:2011ks}
  J.~-F.~Fortin, B.~Grinstein, A.~Stergiou,
  Phys.\ Lett.\  {\bf B704}, 74-80 (2011).
  [arXiv:1106.2540 [hep-th]].

\bibitem{Fortin:2011sz}
  J.~-F.~Fortin, B.~Grinstein, A.~Stergiou,
  [arXiv:1107.3840 [hep-th]].

\bibitem{Nakayama:2009qu}
  Y.~Nakayama,
  arXiv:0907.0227 [hep-th].
\bibitem{Nakayama:2009fe}
  Y.~Nakayama,
  JHEP {\bf 1001}, 030 (2010)
  [arXiv:0909.4297 [hep-th]].
\bibitem{Nakayama:2010wx}
  Y.~Nakayama,
  [arXiv:1009.0491 [hep-th]].

\bibitem{Antoniadis:2011gn}
  I.~Antoniadis, M.~Buican,
  Phys.\ Rev.\  {\bf D83 } (2011)  105011.
  [arXiv:1102.2294 [hep-th]].


\bibitem{Coleman:1970je}
  S.~R.~Coleman and R.~Jackiw,
  Annals Phys.\  {\bf 67} (1971) 552.

\bibitem{Zheng:2011bp}
  S.~Zheng, Y.~Yu,
  [arXiv:1103.3948 [hep-th]].


\bibitem{Freedman:1998rd}
  D.~Z.~Freedman, H.~Osborn,
  Phys.\ Lett.\  {\bf B432}, 353-360 (1998).
  [hep-th/9804101].

\bibitem{Jack:1996qq}
  I.~Jack, D.~R.~T.~Jones, C.~G.~North,
  Nucl.\ Phys.\  {\bf B473}, 308-322 (1996).
  [hep-ph/9603386].

\bibitem{Wallace:1974dy}
  D.~J.~Wallace, R.~K.~P.~Zia,
  Annals Phys.\  {\bf 92 } (1975)  142.

\bibitem{Jack:1990eb}
  I.~Jack, H.~Osborn,
  Nucl.\ Phys.\  {\bf B343}, 647-688 (1990).

\bibitem{Barnes:2004jj}
  E.~Barnes, K.~A.~Intriligator, B.~Wecht, J.~Wright,
  Nucl.\ Phys.\  {\bf B702}, 131-162 (2004).
  [hep-th/0408156].

\bibitem{Komargodski:2011vj}
  Z.~Komargodski, A.~Schwimmer,
  [arXiv:1107.3987 [hep-th]].


\bibitem{Intriligator:2003jj}
  K.~A.~Intriligator, B.~Wecht,
  Nucl.\ Phys.\  {\bf B667}, 183-200 (2003).
  [hep-th/0304128].


\bibitem{Kutasov:2003ux}
  D.~Kutasov,
  [hep-th/0312098].

\bibitem{Erkal:2010sh}
  D.~Erkal, D.~Kutasov,
  [arXiv:1007.2176 [hep-th]].



\bibitem{Nakayama:2011zw}
  Y.~Nakayama,
  [arXiv:1107.2928 [hep-th]].

\end{thebibliography}
\end{document}